%% file: main.tex
\documentclass[11pt]{article}
\PassOptionsToPackage{authoryear}{natbib}
\usepackage[preprint,nonatbib]{neurips_2024}
\usepackage{hyperref} 
\usepackage{authblk} 
\usepackage{booktabs}
\usepackage{graphicx}
\usepackage{amsfonts}
\usepackage{multicol}
\usepackage{tabularx}
\usepackage{enumitem,amsmath,bm,amssymb}
\usepackage{pifont,makecell}
\usepackage{tikz}
\usepackage{CJKutf8}

\usepackage{multirow}
\usepackage{cite}
\usepackage{comment}

\title{JoyVoice: Long-Context Conditioning for Anthropomorphic Multi-Speaker Conversational Synthesis}

\author{
  \bf SpeechTeam, JD
}

\affil{Speech Lab, JD Explore Academy, JD.com Inc.}
\affil{\texttt{joyai\_audio\_llm@jd.com}}
\date{}
\begin{document}

\maketitle

\begin{abstract}
\input{chapters/0_abstract}
\end{abstract}

\input{chapters/1_intro}

\input{chapters/2_joyvoice}

\input{chapters/3_exp}

\input{chapters/4_conclusion}

\input{chapters/5_limitation}

\bibliographystyle{unsrt}
\bibliography{ref}

\appendix

\end{document}

%% file: chapters/0_abstract.tex
Large speech generation models are evolving from single-speaker, short sentence synthesis to multi-speaker, long conversation geneartion. Current long-form speech generation models are predominately constrained to dyadic, turn-based interactions. To address this, we introduce \textbf{JoyVoice}, a novel anthropomorphic foundation model designed for flexible, boundary-free synthesis of up to eight speakers. Unlike conventional cascaded systems, JoyVoice employs a unified \textbf{E2E-Transformer-DiT architecture} that utilizes autoregressive hidden representations directly for diffusion inputs, enabling holistic end-to-end optimization. We further propose a \textbf{MM-Tokenizer} operating at a low bitrate of 12.5 Hz, which integrates multitask semantic and MMSE losses to effectively model both semantic and acoustic information. Additionally, the model incorporates robust text front-end processing via large-scale data perturbation. Experiments show that JoyVoice achieves state-of-the-art results in multilingual generation (Chinese, English, Japanese, Korean) and zero-shot voice cloning. JoyVoice achieves top-tier results on both the Seed-TTS-Eval Benchmark and multi-speaker long-form conversational voice cloning tasks, demonstrating superior audio quality and generalization. It achieves significant improvements in prosodic continuity for long-form speech, rhythm richness in multi-speaker conversations, paralinguistic naturalness, besides superior intelligibility. We encourage readers to listen to the demo at \url{https://jea-speech.github.io/JoyVoice}

%% file: chapters/1_intro.tex
\section{Introduction}

\begin{figure*}[t]
	\centering
	\includegraphics[width=\linewidth]{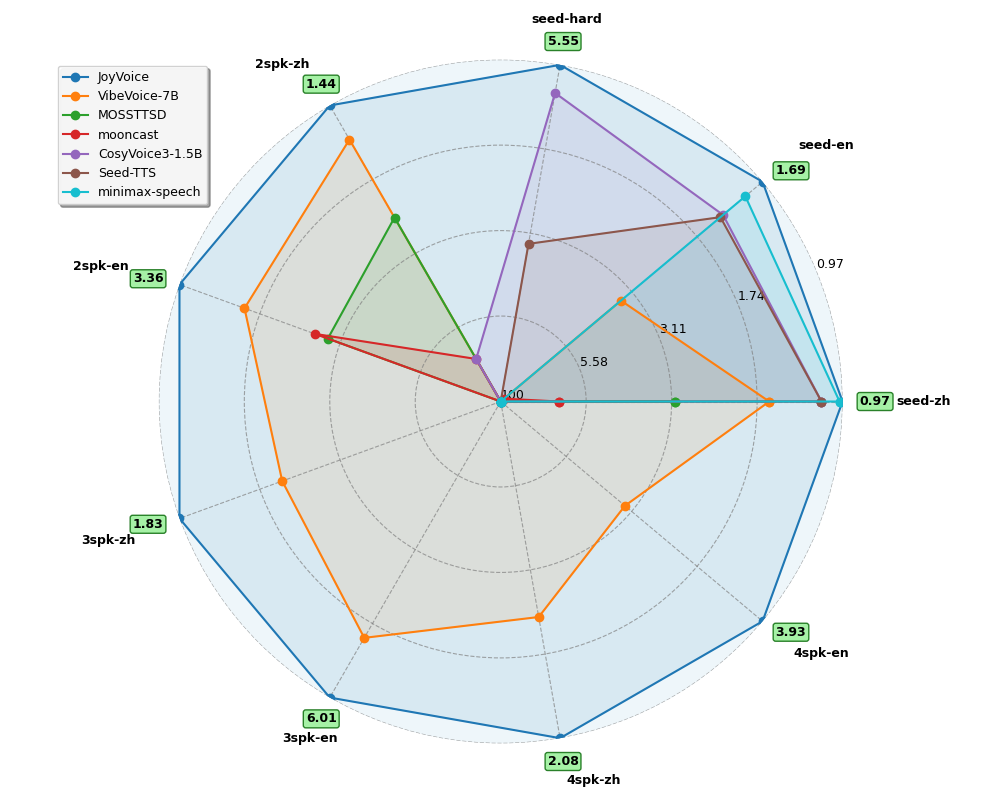}
	\caption{Performance comparison between our JoyVoice and competitive speech generation models in terms of content consistency (in CER/WER) on various benchmarks. The error rates of 100 means that the released models do not support the test tasks.}
	\label{fig:joyvoice_sota}
\end{figure*}

Text-to-speech (TTS) synthesis is the task of generating intelligible and natural synthetic speech from input text messages. In recent years, substantial advancements have been achieved in the field of TTS, particularly in zero-shot voice cloning, driven by the emergence of sophisticated generative models such as large language models (LLMs) and diffusion models, coupled with expanding training data scales and model sizes. Compared to conventional unit selection, concatenative, and statistical parametric approaches, modern TTS models can now achieve results that are often indistinguishable from genuine human speech.

The current mainstream paradigms in TTS are generally categorized into autoregressive and non-autoregressive methods. The autoregressive approach typically utilizes a decoder-only transformer to predict discrete speech tokens from text \cite{vall-e,ye2025llasa}. Conversely, non-autoregressive models, such as diffusion models, directly predict acoustic features from the text input \cite{chen2025f5tts, zhu2025zipvoice}. While autoregressive generation offers advantages in streaming capability and reduced latency, non-autoregressive models generally boast a lower real-time factor (RTF), making each suitable for different application requirements. Recently, several two-stage TTS models \cite{anastassiou2024seed,du2024cosyvoice,du2024cosyvoice2,du2025cosyvoice3,wang2025spark,guo2024fireredtts,zhou2025indextts2,zhang2025minimax} have been proposed. These models combine an autoregressive LLM to generate speech-derived semantic tokens from text, with a non-autoregressive model (like a diffusion model or its variants, such as flow-matching models \cite{lipman2022flowmatching} and Diffusion Transformers (DiTs) \cite{peebles2023dit}) to synthesize acoustic features from these predicted tokens.

These discrete semantic tokens serve as a crucial intermediate representation between text and acoustic information. However, the inherent quantization prevents gradient backpropagation from the acoustic synthesis stage (diffusion models) to the token prediction stage (LLMs), necessitating separate and independent training of these components. Furthermore, common speech tokenizers \cite{du2024cosyvoice, du2024cosyvoice2} maintain a 1:2 semantic-to-acoustic token ratio, resulting in excessively long output sequences for the LLM, thereby reducing inference efficiency. Longer sequences also increase the risk of LLM-induced generation artifacts, commonly known as hallucination.

In contrast to the aforementioned two-stage models, a more recent trend involves innovative one-stage frameworks based on continuous features, often leveraging variational autoencoder (VAE) representations \cite{peng2025vibevoice,liu2024ardit}. In this paradigm, the hidden states of the LLMs condition local DiTs to generate continuous representations. This approach seamlessly integrates the text-to-token and token-to-acoustic stages, enabling joint optimization of all model components. By avoiding quantization, the continuous VAE representations preserve the fidelity of the original information and can be efficiently compressed into patches before being input to the LLMs \cite{jia2025ditar}. This compression significantly reduces the effective frame rate for the LLM component (to as low as 5–10 Hz), enhancing computational efficiency and the capability for generating long speech.

Nevertheless, these one-stage models face their own set of challenges. The limited receptive field of local DiT components can compromise timbre similarity. Moreover, if the prompt speech contains noise, this contamination can be propagated to the LLM via the continuous representations, leading to instability during synthesis. Consequently, generating high-quality, expressive speech for complex texts and challenging scenarios remains constrained. Furthermore, the latest large-scale TTS models are beginning to shift from single-speaker, short-form audio generation toward multi-speaker, long-form audio generation. However, current work primarily focuses on two-person turn-by-turn synthesis~\cite{ju2025mooncast,moss2025ttsd,zhu2025zipvoicedialognonautoregressivespokendialogue}. VibeVoice~\cite{peng2025vibevoice} is one of the few large-scale TTS models capable of supporting long-form audio generation with more than two speakers. Yet, ensuring the stability and naturalness of multi-speaker, long-context, multi-turn conversational audio synthesis in a zero-shot TTS setting remains an open and highly challenging problem.

To address these limitations and, fundamentally, to enable scalable and stable synthesis of multi-speaker, long-form audio with high naturalness and speaker fidelity, we propose JoyVoice, an efficient and scalable zero-shot TTS synthesizer. JoyVoice natively supports long-context conditioning, as well as multi-speaker and multi-turn conversational synthesis. Similar to other one-stage TTS models \cite{jia2025ditar}, JoyVoice employs an integrated architecture where the LLM's hidden states condition the DiTs, allowing for seamless joint optimization. A key difference, however, is our utilization of a global causal DiT model, which is specifically designed to model long-context sequences, thus enhancing the speaker similarity of the synthesized speech. Simultaneously, we strategically reincorporate discrete speech tokens, akin to those in previous two-stage models like CosyVoice, to bolster system stability. We deploy a dynamic chunk-wise attention strategy on the DiTs to unify streaming and non-streaming synthesis within a single framework, allowing flexible adaptation to various latency requirements. Motivated by CosyVoice 3 \cite{du2025cosyvoice3}, we introduce a supervised speech semantic tokenizer (MM-tokenizer) optimized via a multi-task learning approach, achieving a highly efficient token rate compressed to 12.5 Hz. This high compression rate is vital for efficient long-context modeling and long-form audio generation. Furthermore, by employing a curriculum learning strategy, JoyVoice effectively supports long-context, multi-speaker, and multi-turn conversation generation. Both subjective and objective evaluation results demonstrate that the proposed JoyVoice achieves superior prosody naturalness, content consistency, and speaker similarity compared to state-of-the-art TTS synthesizers. Additionally, reinforcement learning is introduced during the post-training phase of JoyVoice to mitigate pronunciation errors and unnatural outputs, further enhancing model stability.

Our contributions are: 
\begin{itemize}[leftmargin=*] 
\item Integrated End-to-End Architecture: We propose a unified framework where AR-Transformer hidden states directly condition a global causal DiT, ensuring seamless optimization and superior multi-speaker synthesis. 
\item High-Efficiency MM-Tokenizer: We introduce a tokenizer operating at 25Hz \& 12.5 Hz that balances semantic understanding and acoustic reconstruction through multi-task losses. 
\item Robustness without Normalization: By leveraging large-scale data perturbation and synthetic generation, JoyVoice reduces dependency on text normalization modules while maximizing stability. 
\item SOTA Performance: JoyVoice achieves state-of-the-art results on the Seed-TTS-Eval Benchmark and complex conversational voice cloning tasks, surpassing existing baselines in prosody, consistency, and similarity. 
\end{itemize}

%% file: chapters/2_joyvoice.tex
\section{JoyVoice}

\begin{figure*}[t]
	\centering
	\includegraphics[width=\linewidth]{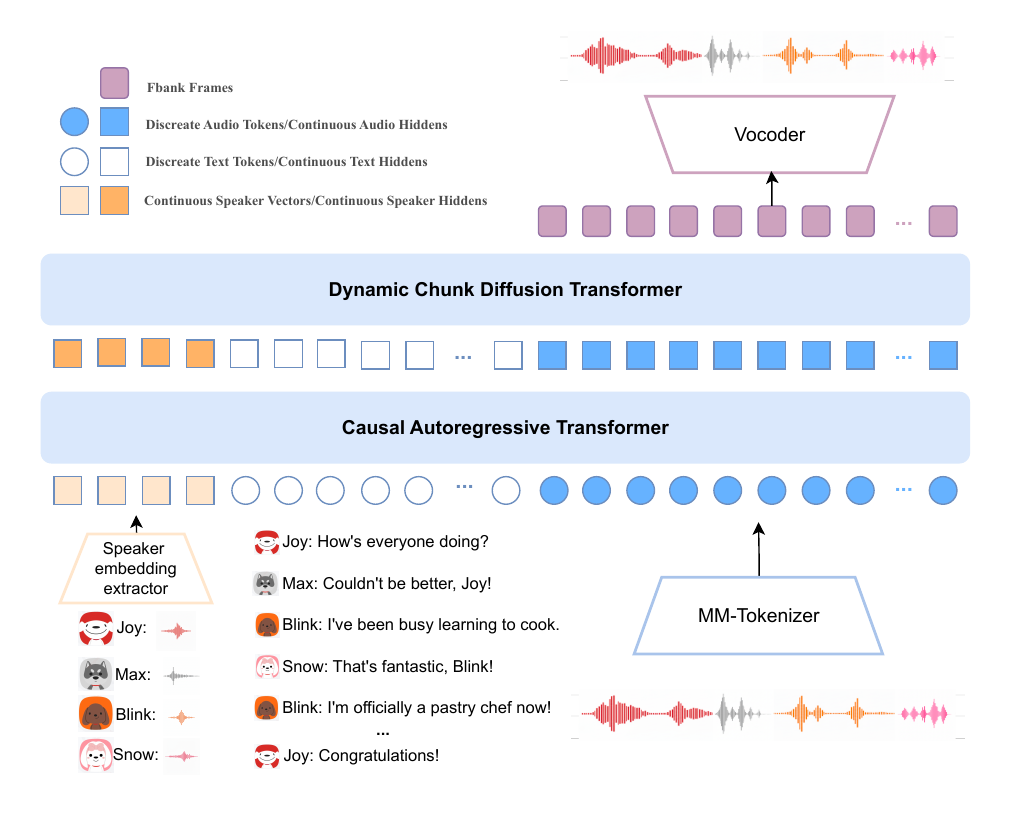}
	\caption{An overview of JoyVoice. It Consists of several key components: 1) Causal Autoregressive Transformer: This module takes the system prompt plus text with speaker tags as input, and predicts the discrete tokens generated by the MM-Tokenizer. 2) Dynamic Chunk Diffusion Transformer: Using the hidden representations from the Causal Autoregressive Transformer as input, this component predicts the mel-spectrogram output. 3) MM-Tokenizer and Vocoder: These modules are responsible for converting audio into discrete token representations and reconstructing the mel-spectrogram back to waveform audio, respectively.}
	\label{fig:joyvoice}
\end{figure*}

\subsection{E2E-Transformer-DiT Architecture}

Previous coarse-to-fine TTS systems~\cite{anastassiou2024seed,du2024cosyvoice,du2024cosyvoice2,du2025cosyvoice3,wang2025spark,guo2024fireredtts,zhou2025indextts2,zhang2025minimax} typically employ a two-stage pipeline: an autoregressive transformer acoustic model (AM) for discrete token prediction followed by a separate flow matching (FM) model for token-to-mel acoustic detail recovery. While this decoupled approach offers modularity, it suffers from several inherent limitations: (1) Information bottleneck: the discrete tokens generated by AM may lose fine-grained prosodic and acoustic information critical for high-fidelity synthesis; (2) Optimization inconsistency: the AM is trained to maximize token prediction likelihood without considering the downstream FM's reconstruction quality, leading to suboptimal overall performance; (3) Error propagation: prediction errors from the AM directly impact FM's conditioning, with no mechanism for the FM to provide feedback to improve AM's output quality.

We propose a joint training framework that unifies the autoregressive acoustic modeling and flow-based acoustic refinement stages. By allowing gradient flow from the FM loss back to the AM, our approach enables the acoustic model to learn representations that are not only optimal for discrete token prediction but also maximally informative for continuous acoustic detail recovery. This joint optimization paradigm naturally encourages the AM to preserve acoustic information that discrete tokens alone cannot capture, while simultaneously enabling the FM to guide the AM toward generating more reconstruction-friendly hidden representations.

Our proposed architecture, illustrated in Figure \ref{fig:joyvoice}, consists of two core components: (1) a Qwen-based~\cite{bai2023qwen} autoregressive acoustic model that predicts these tokens from text, and (2) a flow matching model that recovers fine-grained acoustic details. The key innovation lies in our end-to-end joint training strategy, where the AM's hidden states are directly fed as conditioning signals to the FM, enabling bidirectional gradient propagation.

The joint training framework exhibits several key advantages:
\begin{itemize}[leftmargin=*]
\item \textbf{Instruction-aware acoustic refinement for Flow Matching}:
Unlike previous cascaded systems where only phonetic content, in the form of speech tokens, from the AM conditions the FM, our architecture enables the complete instruction information—including style directives, speaker characteristics, and prosodic controls—to be directly propagated to the FM. This allows the flow matching model to leverage high-level semantic guidance for sculpting acoustic details beyond mere pronunciation, resulting in more controllable and context-aware synthesis. 

\item \textbf{Multi-speaker long-form audio modeling capability}: This end-to-end optimization mechanism is crucial for achieving multi-speaker long-form audio generation in JoyVoice. Since the discretized semantic tokens contain limited speaker-related information, conventional two-stage models can only perform turn-by-turn prediction of semantic tokens while requiring prior knowledge of speaker boundaries in the semantic tokens. In contrast, JoyVoice directly employs hidden continuous representations rich in information as input to the foundation model and optimizes the entire system in an end-to-end manner, enabling the model to automatically learn discriminative information for both speaker characteristics and semantic content. Consequently, without requiring explicit knowledge of speaker acoustic boundaries, JoyVoice can realistically generate long-form audio with 1 to 8 speakers.

\item \textbf{Improved intelligibility through end-to-end optimization}: Experimental results demonstrate that our jointly trained model achieves significantly lower Word Error Rate (WER) compared to its cascaded counterpart, indicating that the bidirectional gradient flow enables the AM to generate representations that are inherently more conducive to accurate acoustic reconstruction. 

\item \textbf{Robustness to tokenizer compression}: When reducing the tokenizer's frame rate from 25 Hz to 12.5 Hz—a fifty percent compression that substantially increases the information bottleneck—the cascaded baseline suffers notable performance degradation. In contrast, our end-to-end framework maintains performance comparable to the 25 Hz setting, demonstrating that joint optimization effectively compensates for information loss in discrete tokens by allowing the AM to encode complementary continuous information in its hidden representations that the FM can exploit for reconstruction.
\end{itemize}

The overall training loss is formulated as:
\begin{equation}
\mathcal{L} = \mathcal{L}_{\text{AM}} + \lambda \cdot \mathcal{L}_{\text{FM}}
\end{equation}

where $\lambda$ is a balancing hyperparameter that controls the relative contribution of each component.

\textbf{Autoregressive Acoustic Model Loss}: The AM is trained with a standard next-token prediction objective using cross-entropy loss:
\begin{equation}
\mathcal{L}_{\text{AM}} = -\sum_{t=1}^{T} \log p(s_t \mid \mathcal{I}_{<t}; \theta_{\text{AM}})
\end{equation}
where $s_t$ is the ground-truth discrete speech token at position $t$, and $\mathcal{I}_{<t}$ denotes context including all speaker information, all text tokens, and previously generated speech tokens $\{s_i \mid i < t\}$. The detailed formulation of $\mathcal{I}$ will be presented in the next section.

\textbf{Flow Matching Model Loss}: The FM is trained to recover continuous acoustic features from the AM's hidden representations. Following the standard flow matching formulation, we optimize:
\begin{equation}
\mathcal{L}_{\text{FM}} = \mathbb{E}_{t \sim \mathcal{U}(0,1), x_0, x_1} \left[ | v_\theta(x_t, t, \mathbf{h}_{\text{AM}}) - (x_1 - x_0) |^2 \right]
\end{equation}
where $x_0 \sim \mathcal{N}(0, I)$ is sampled from a standard Gaussian prior, $x_1$ represents the ground-truth acoustic features (e.g., mel-spectrograms), $x_t = (1-t)x_0 + tx_1$ is the linear interpolation, $v_\theta$ is the velocity field predicted by the FM, and $\mathbf{h}_{\text{AM}}$ denotes the hidden states extracted from the AM that serve as conditioning signals.
For the gradient flow mechanism, the key innovation lies in conditioning the FM on $\mathbf{h}_{\text{AM}}$ rather than solely on discrete tokens. During backpropagation, gradients from $\mathcal{L}_{\text{FM}}$ flow through $\mathbf{h}_{\text{AM}}$ back to the AM parameters $\theta{_\text{AM}}$:
\begin{equation}
\frac{\partial \mathcal{L}}{\partial \theta{_\text{AM}}} = \frac{\partial \mathcal{L}_{\text{AM}}}{\partial \theta{_\text{AM}}} + \lambda \cdot \frac{\partial \mathcal{L}_{\text{FM}}}{\partial \mathbf{h}_{\text{AM}}} \cdot \frac{\partial \mathbf{h}_{\text{AM}}}{\partial \theta{_\text{AM}}}
\end{equation}
This bidirectional optimization enables the AM to learn representations that simultaneously optimize discrete token prediction accuracy and continuous acoustic reconstruction quality, effectively addressing the information bottleneck inherent in cascaded architectures.

\subsection{Multi-speaker Multi-turn Modeling with Long Continuous Speech Context}
Previous pipelines for multi-speaker, multi-turn dialogue synthesis adopt a modular, utterance-level strategy~\cite{peng2025vibevoice,ju2025mooncast}. In these frameworks, the conversation is processed as a sequence of isolated text-speech pairs, where each text input for turn \( i \) is independently converted into its corresponding speech output. Although this formulation provides explicit utterance boundaries, it introduces artificial segmentation that can disrupt the natural flow of conversational dynamics and requires precise forced alignment during training.

In contrast, our approach introduces a unified sequence formulation that maintains the natural continuity of multi-turn dialogues without relying on utterance-level segmentation. Formally, for a conversation involving \( N \) speakers and \( M \) turns, we construct the input sequence \( \mathcal{I} \) as follows:

\begin{equation}
\mathcal{I} = \big[ P;\ T;\ S \big]
\end{equation}

where
\begin{equation}
\begin{aligned}
P &= \big[ \text{spk}_0, \mathbf{e}_0, \text{spk}_1, \mathbf{e}_1, \ldots, \text{spk}_{N-1}, \mathbf{e}_{N-1} \big], \\
T &= \big[ \text{spk}_{i_0}, \mathbf{t}_0, \text{spk}_{i_1}, \mathbf{t}_1, \ldots, \text{spk}_{i_{M-1}}, \mathbf{t}_{M-1} \big], \\
S &= \big[ s_1, s_2, \ldots, s_T \big].
\end{aligned}
\end{equation}

Here, \( \text{spk}_k \) denotes the speaker tag for the \( k \)-th speaker, \( \mathbf{e}_k \) is the corresponding speaker embedding, \( \mathbf{t}_j \) represents the text token sequence of the \( j \)-th turn (with speaker index \( i_j \in \{0, 1, \ldots, N-1\} \) and variable sequence length), and \( s_i \) denotes a single speech token. Note that \( S \) represents the speech token sequence spanning the entire multi-turn dialogue without segmentation.

By adopting implicit turn-level correspondence learning without explicit speech segmentation, the model associates speaker tags with their corresponding embeddings, text content, and speech token spans through attention mechanisms, enabling flexible speaker-content binding. Maintaining the entire dialogue as a continuous sequence preserves the conversational context and allows the model to capture cross-turn dependencies, turn-taking dynamics, and long-range prosodic patterns that would be fragmented in segmented approaches. Experimental validation demonstrates that modern Transformer attention mechanisms can accurately establish correspondences between unsegmented speech tokens and their associated text and speaker identities using only speaker tags.

\subsection{MM-Tokenizer}
\label{sec:mm_tokenizer}
\begin{figure*}[thb]
	\centering
	\includegraphics[width=\linewidth]{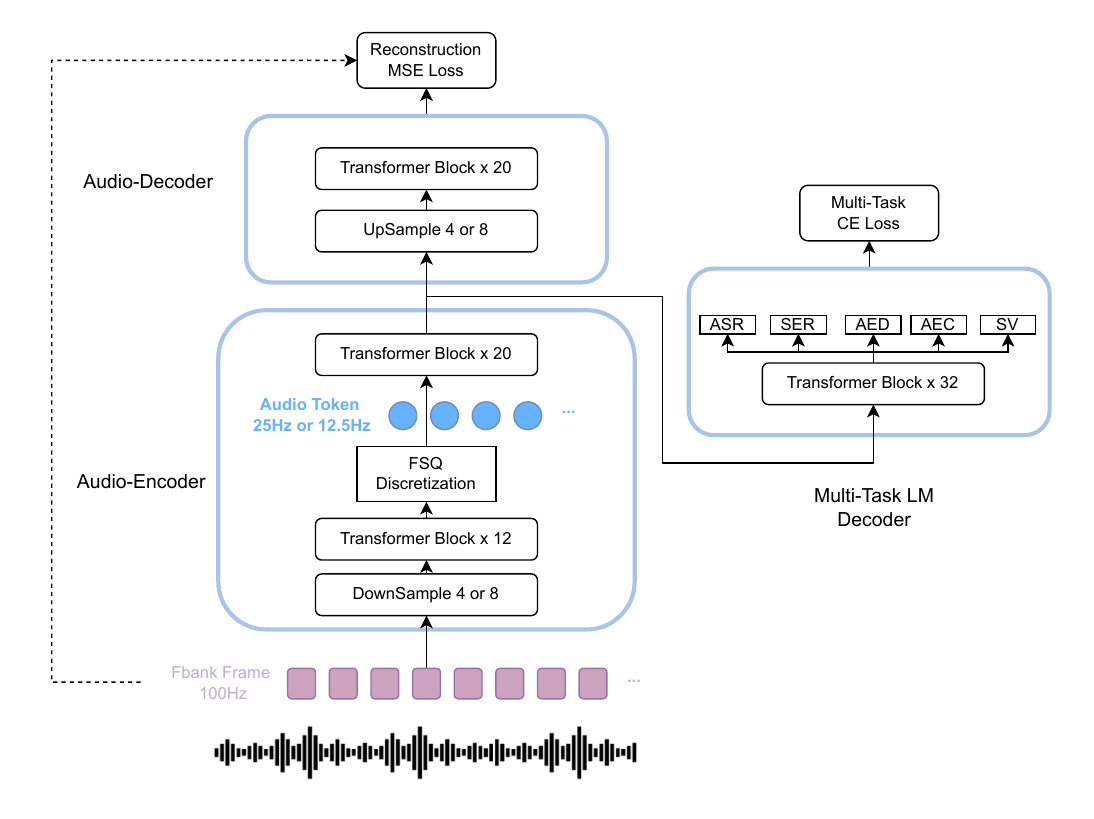}
	\caption{Illustration of JoyVoice MM-Tokenizer framework.}
	\label{fig:MM-tokenizer}
\end{figure*}

As illustrated in Figure \ref{fig:MM-tokenizer}, JoyVoice employs an MM-Tokenizer based on supervised multi-task training semantic Token scheme similar to CosyVoice 3~\cite{du2025cosyvoice3}. We adopt the Whisper-large-v3~\cite{radford2023robust} ASR model as the pretrained model, and integrated a Finite Scalar Quantization (FSQ) ~\cite{mentzer2023finite} module into the linear layer of the 12th Transformer block in the encoder. The model was trained using a supervised multi-task learning approach, which encompasses not only audio understanding tasks—such as automatic speech recognition (ASR), speech emotion recognition (SER), audio event detection (AED), audio event captioning (AEC), speaker verification (SV), age detection (AD), and gender classification (GC). We augment the standard Whisper-large-v3 vocabulary with specialized discrete tokens to represent various audio understanding tasks, such as $\texttt{<|AED|>}$ for audio event detection (AED), etc. This allows the decoder to perform sequence-to-sequence generation for multiple tasks simultaneously. An external audio decoder module is designed to reconstruct the original Mel-spectrogram from the encoder output, which ensures that the learned discrete representations are more generalizable and robust, capturing both semantic and acoustic information. The final objective function for training the MM-Tokenizer is the weighted combination of the two primary loss components:
$$
\min_{\Theta} \mathcal{L}_{\text{total}} = \mathcal{L}_{\text{semantic}} + \beta \mathcal{L}_{\text{recon}}
$$
, where $\beta$ is a scalar hyperparameter controlling the trade-off between semantic performance and acoustic reconstruction quality.

Notably, in our experiments, we explored discrete representations at two different token rates: 25 Hz and 12.5 Hz. To achieve these rates, we modified the original Whisper-large-v3 encoder by incorporating additional CNN layers for 4× and 8× downsampling, respectively. Subsequent experiments will compare the performance of the two token rates. A noteworthy finding is that our proposed joint modeling approach reduces dependency on the token rate: the 12.5 Hz representation performs comparably to the 25 Hz representation across both subjective and objective metrics.

\subsection{Dynamic Chunk Flow Matching}
To enable streaming synthesis, a causal chunk-wise Flow Matching model is adopted, which divides LLM hidden outputs into small chunks.
A chunk-wise causal mask motivated by other steaming systhesis models \cite{du2024cosyvoice2,ju2025mooncast} is deployed on self-attention modules in each Transformer layer, which can guarantee each current chunk only accesses the historical context of previous chunks. For the convenience of inference process, we conduct a random chunk size on each training step. Since the fixed chunk size is eliminated in the training stage, arbitrary chunk sizes can be flexibly adopted at inference time, enabling the flexibility to support a wide range of application scenarios and different latency constraints.


\subsection{TTS-Frontend-Free Optimization}
Conventional text-to-speech (TTS) systems typically incorporate a front-end text preprocessing module responsible for normalizing input text—handling elements such as punctuation and text normalization (TN). This is particularly critical in Chinese TTS systems, which must also address rare characters and polyphonic words. In JoyVoice, we aim to emulate the robustness of text-based foundation models, enabling the system to process diverse input text without relying on a dedicated TTS frontend. Consequently, none of the experiments or demo samples presented in this work utilize any preprocessing module. To achieve this, in addition to initializing the model with a Qwen text foundation model, we performed targeted data augmentation covering TN/ITN, polyphonic characters, and rare characters.

\subsection{Curriculum Learning}
We adopt a two-stage curriculum learning strategy to enhance the training stability of JoyVoice for modeling multi-speaker long-form audio. In the first stage, the model is pre-trained using large-scale single-speaker short audio segments, with each clip limited to a maximum length of one minute. This pre-training phase yields a high-quality base model capable of synthesizing single-speaker speech effectively. In the second stage, we combine the full dataset—comprising long single-speaker audio, multi-speaker long-form audio, and the original short single-speaker data—to continue fine-tuning the model. The pre-trained JoyVoice model serves as the initialization checkpoint for this phase. During fine-tuning, the audio duration is extended up to five minutes and the number of speakers is increased to a maximum of eight. Through this two-stage curriculum learning approach, JoyVoice not only achieves strong performance in zero-shot single-speaker speech synthesis but also delivers high-quality zero-shot generation for multi-speaker long-form audio.

\subsection{Reinforcement Learning}

\subsubsection{Preference Alignment}
The alignment of language models with human preferences is commonly framed as a reinforcement learning (RL) problem. Given an input prompt $x$ and the model's response $y$, the objective is to learn a policy $\pi_\theta$ that maximizes the expected reward while maintaining proximity to a reference policy $\pi_{\text{ref}}$. This is formalized as follows.
\begin{equation}
\max_{\pi_\theta} \mathbb{E}_{y \sim \pi_\theta(y|x)} [r(x, y)] - \beta D_{\text{KL}} (\pi_\theta(y|x) \| \pi_{\text{ref}}(y|x))
\end{equation}
where $r(x, y)$ represents the reward function, and the KL-divergence term, controlled by the hyperparameter $\beta$, constrains the policy deviation. Since the true reward function is typically unknown, it is estimated from human preference data consisting of tuples $(x, y_w, y_l)$, where $y_w$ denotes the preferred response and $y_l$ the disfavored one. The reward function can be approximated through maximum likelihood estimation:
\begin{equation}
\hat{r} \in \arg \min_r \mathbb{E}_{(x,y_w,y_l)} [-\log \sigma (r(x, y_w) - r(x, y_l))]
\end{equation}
with $\sigma$ being the sigmoid function.

\subsubsection{Direct Preference Optimization}
Direct Preference Optimization (DPO) \cite{rafailov2023direct} have enabled the solving of this optimization problem in a closed form without explicit reward modeling, which reparameterizes the reward function using the optimal solution to the KL-constrained objective:
\begin{equation}
r(x, y) = \beta \log \left( \frac{\pi_\theta(y|x)}{\pi_{\text{ref}}(y|x)} \right) + \beta \log Z(x)
\end{equation}
Under the Bradley-Terry model, the preference probability is given by:
\begin{equation}
P(y_w \succ y_l|x) = \sigma \left( \beta \log \frac{\pi_\theta(y_w|x)\pi_{\text{ref}}(y_l|x)}{\pi_\theta(y_l|x)\pi_{\text{ref}}(y_w|x)} \right)
\end{equation}
The DPO objective function directly maximizes this likelihood:
\begin{equation}
\mathcal{L}_{\text{DPO}} = \mathbb{E}_{(y_w,y_l,x)} \left[ -\log \sigma \left( \beta \log \frac{\pi_\theta(y_w|x)\pi_{\text{ref}}(y_l|x)}{\pi_\theta(y_l|x)\pi_{\text{ref}}(y_w|x)} \right) \right]
\end{equation}

\subsubsection{APO: Acoustic Preference Optimization for Text-to-Speech}
We introduce Acoustic Preference Optimization (APO), a novel approach specifically designed for text-to-speech (TTS) systems that extends preference optimization to the acoustic token level. In modern neural codec-based TTS systems, speech waveforms are converted into discrete acoustic token sequences through neural audio codecs. Although DPO-based methods cannot be applied directly to optimize final waveform outputs, APO enables fine-grained optimization of individual acoustic token predictions within generated speech sequences. This approach is particularly valuable for TTS, where token-level quality variations significantly impact overall speech naturalness and intelligibility.

\paragraph{Acoustic Token-Level Preference Construction}
For a given text input $x$, we generate $N$ candidate acoustic token sequences $\{s_1, s_2, \ldots, s_N\}$, where each sequence $s_i = \{t_{i,1}, t_{i,2}, \ldots, t_{i,n}\}$ corresponds to a potential speech output. We evaluate each sequence using the character error rate (CER) metric. The preference sets are constructed on the basis of these evaluations:

\begin{equation}
\mathcal{S}_{\text{chosen}} = \{s_i \in \{s_1, \ldots, s_N\} | \text{CER}(s_i) = 0\}
\end{equation}
\begin{equation}
\mathcal{S}_{\text{rejected}} = \{s_j \in \{s_1, \ldots, s_N\} | \text{CER}(s_j) > 0\}
\end{equation}

where $\mathcal{S}_{\text{chosen}}$ contains all sequences with perfect character accuracy (CER = 0), and $\mathcal{S}_{\text{rejected}}$ contains sequences with transcription errors. During training, we form preference pairs by creating all possible combinations between sequences from $\mathcal{S}_{\text{chosen}}$ and $\mathcal{S}_{\text{rejected}}$. Formally, the set of all preference pairs, denoted as $\mathcal{P}$, is constructed as:
\begin{equation}
\mathcal{P} = \mathcal{S}_{\text{chosen}} \times \mathcal{S}_{\text{rejected}} = \{(s_{\text{chosen}}, s_{\text{rejected}}) \mid s_{\text{chosen}} \in \mathcal{S}_{\text{chosen}}, s_{\text{rejected}} \in \mathcal{S}_{\text{rejected}}\}
\end{equation}
This approach ensures that the model learns to distinguish between perfectly accurate token sequences and those with errors, directly optimizing for token-level correctness in the generated speech.

%% file: chapters/3_exp.tex
\section{Multi-Speaker Voice Clone Evaluation Benchmark}
With the rapid development of speech generation models, existing evaluation benchmarks no longer meet the model assessment requirements, especially for multi-speaker long-form audio modeling. 
Firstly, current multi-speaker dialogue evaluation benchmarks such as zipvoice-dialog~\cite{zhu2025zipvoicedialognonautoregressivespokendialogue} are limited in the number of speakers—typically involving only two participants, which restricts diversity.
In practical domains such as audiobooks, podcasts, and live streaming, it is common to observe multi-speaker conversations involving two to four or more participants.  This renders existing benchmarks, primarily designed for two-speaker scenarios, inadequate for assessing model performance in these diverse and complex settings.
Secondly, the dialogue scripts in existing benchmarks predominantly comprise short segments under one minute, thereby lacking the contextual coherence required for evaluating sustained conversational interactions.
Finally,  traditional benchmarks prioritizes character accuracy and single-utterance speaker similarity, is inadequate for multi-spkeaker long dialogues. It fails to account for the critical need to jointly assess timbre stability across turns, character accuracy, and overall coherence in long-form audio—precisely the challenges exacerbated by frequent speaker changes.

To better evaluate JoyVoice, we establish a benchmark for multi-speaker multi-turn long-form Chinese and English audio, named JoyVoice-MSMT-eval. The details are as follows:

\begin{itemize}[leftmargin=*]
\item \textbf{Multi-Speaker:}: The benchmark covers dialogues involving 2 to 4 speakers. Both the Chinese and English subsets include 50 two-speaker dialogues, 50 three-speaker dialogues, and 25 four-speaker dialogues. Source audio is sampled from the SEED-TTS-eval. To ensure that all source audio (prompt audio) segments in a long dialogue come from distinct speakers, we performed both speaker similarity verification and manual validation.
\item \textbf{Long-Form Audio}: The benchmark includes scripts corresponding to 1- to 5-minute long audio segments, designed to assess the long-form synthesis capability of multi-speaker dialogue speech generation models. All long-form scripts were generated by a large language model, with each script meticulously reviewed and approved by human annotators to ensure high quality and contextual continuity.
\item \textbf{Evaluation Metric}: Since character error rate (CER) alone is insufficient to capture timbre variations during frequent speaker turns, we introduce concatenated minimum-permutation word error rate (cpWER)\footnote{The cpWER is computed as follows: (i) concatenate all reference transcriptions in chronological order for each speaker. (ii) concatenate all hypothe-sis transcriptions in chronological order for each detected speaker. (iii) compute the WER between the concatenated references and concatenated hypotheses for all possible speaker permutations, and pick the lowest WER among them.}~\cite{Yu2022ACS,kanda2021comparative} as an objective evaluation metric, which is affected by both the speech recognition and speaker diarization results. 

\end{itemize}

\section{Experimental Results}

\subsection{Training Data}

\begin{figure*}[t]
	\centering
	\includegraphics[width=\linewidth]{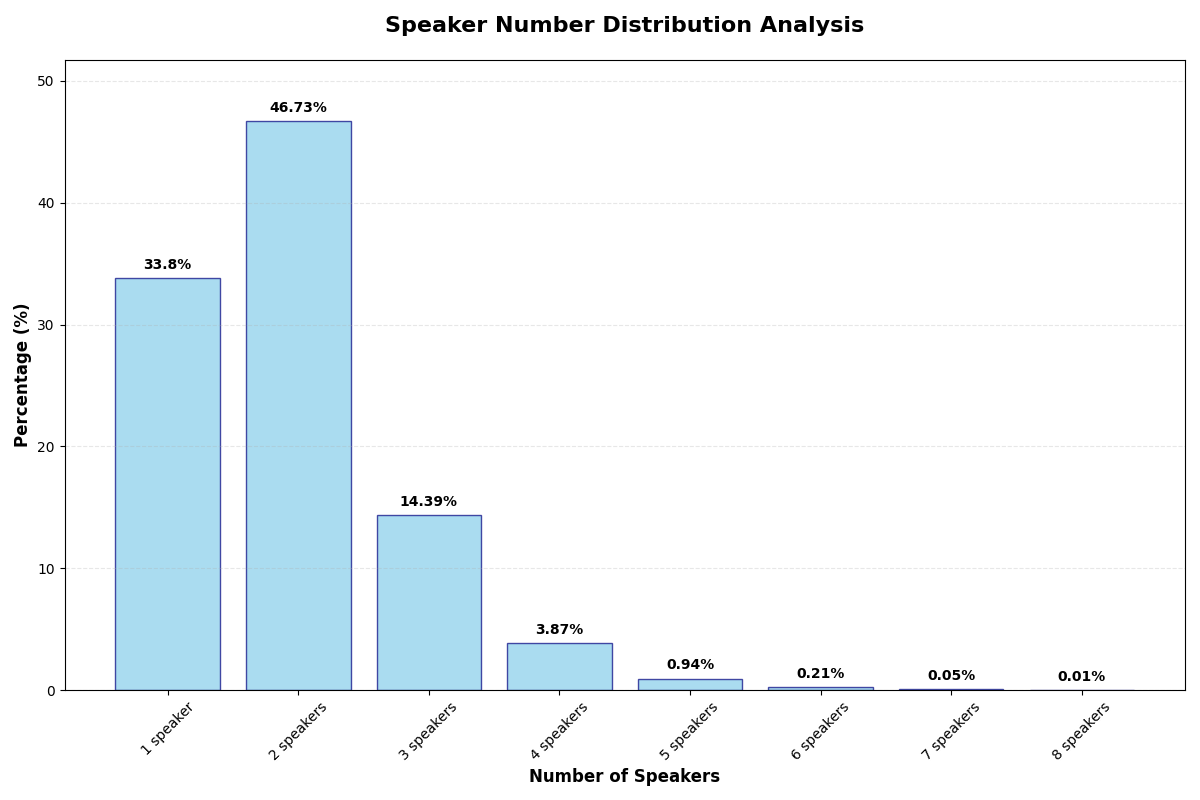}
	\caption{Speaker Distribution in JoyVoice's Long-Form Audio Training Data.}
	\label{fig:spk_dist}
\end{figure*}
We have developed a refined multilingual data production pipeline for constructing the JoyVoice training dataset. 
The linguistic coverage includes Chinese (encompassing a minor proportion of Cantonese and other Chinese dialects), English, Japanese, and Korean, with Chinese and English collectively accounting for more than 90\% of the total dataset. Our automated pipeline executes the following processing stages: audio normalization, quality assessment, enhancement and separation, voice activity detection (VAD) timestamp annotation, speaker diarization, multiple speech recognition systems for text transcription, punctuation restoration. We holistically leverage metadata derived from preceding stages to filter and select high-quality samples.

We constructed two distinct training data formats. \textbf{Short-utterance data (BaseData)}: Audio clips typically under 30 seconds, strictly containing single-speaker utterances. \textbf{Long-form audio data (LongData)}: Audio segments ranging from 30 seconds to 5 minutes, containing 1--8 speakers. 
The speaker distribution within the LongData subset is presented in Figure~\ref{fig:spk_dist}.

\subsection{Model Configuration}
\label{subsec:model_config}

In addition to the proposed JoyVoice, which is based on the E2E-Transformer-DiT architecture, we also developed a two-stage cascade system to serve as a comparative baseline.
For clarity throughout the experimental section, we denote the end-to-end system as \emph{JoyVoice} and the two-stage variant as \emph{JoyVoice-Cascade}.

Both architectures initialize the Autoregressive (AR) Transformer using the Qwen2.5-0.5B LLM~\cite{qwen2.5}. For the Flow Matching component, we employ the Diffusion Transformer (DiT) architecture from F5-TTS~\cite{chen2025f5tts}, comprising approximately 300 million parameters. It is worth noting that JoyVoice-Cascade is trained exclusively on single-speaker data and therefore does not support multi-speaker speech synthesis.

The models are trained using the AdamW optimizer with a linear warmup and cosine annealing schedule. Specifically, the warmup steps are set to $10,000$, with a peak learning rate of $1 \times 10^{-4}$.

\subsection{Evaluation on MM-Tokenizer}
\begin{table}[htb]
\centering
\setlength\tabcolsep{10pt} 
\renewcommand{\arraystretch}{1.1} 
\begin{tabular}{l c c c}
\toprule
\textbf{Benchmark} & \textbf{Whisper-large-v3} & \textbf{MM-Tok (25Hz)} & \textbf{MM-Tok (12.5Hz)} \\
\midrule
\multicolumn{4}{l}{\textit{\textbf{ASR Tasks} (WER/CER) $\downarrow$}} \\
\midrule
AISHELL-1 & 5.14 & \textbf{2.53} & 2.60 \\
AISHELL-2 & 4.96 & \textbf{3.56} & 3.84 \\
{LibriSpeech-clean} & \textbf{1.80} & 2.35 & 2.67 \\
{LibriSpeech-other} & \textbf{3.60} & 5.24 & 6.62 \\
{Fleurs-zh} & 7.70 & \textbf{4.25} & 4.72 \\
{Fleurs-en} & \textbf{4.25} & 8.00 & 9.52 \\
\midrule
\multicolumn{4}{l}{\textit{\textbf{Audio Understanding} (Accuracy) $\uparrow$}} \\
\midrule
ESD (SER) & \multicolumn{1}{c}{-} & \textbf{0.886} & 0.850 \\
FSDKaggle (AED) & \multicolumn{1}{c}{-} & \textbf{0.548} & 0.530 \\
\bottomrule
\end{tabular}
\vspace{2mm}
\caption{\textbf{Comparison on Audio Understanding Tasks.} We compare the proposed MM-Tokenizer (25Hz and 12.5Hz) against the Whisper-large-v3 baseline. $\downarrow$ indicates lower is better (WER/CER), and $\uparrow$ indicates higher is better (Accuracy). The best results are highlighted in \textbf{bold}.}
\label{tab:MM_tokenizer_upstream}
\end{table}

\begin{table}[htb]
\centering
\renewcommand{\arraystretch}{1.2} 
\setlength\tabcolsep{5pt} 

\begin{tabular}{lccccccc} 
\toprule
\multirow{2.5}{*}{\textbf{Model}} & \multirow{2.5}{*}{\textbf{Arch.}} & \multicolumn{2}{c}{\textbf{\textit{test-zh}}} & \multicolumn{2}{c}{\textbf{\textit{test-en}}} & \multicolumn{2}{c}{\textbf{\textit{test-hard}}} \\
\cmidrule(lr){3-4} \cmidrule(lr){5-6} \cmidrule(lr){7-8} 
 & & \textbf{CER}$\downarrow$ & \textbf{SS}$\uparrow$ & \textbf{WER}$\downarrow$ & \textbf{SS}$\uparrow$ & \textbf{CER}$\downarrow$ & \textbf{SS}$\uparrow$ \\
\midrule

\multicolumn{8}{l}{\textit{Frame Rate: 25Hz}} \\
\hspace{3mm} $S^3$-Tokenizer~\cite{du2024cosyvoice} & Cascade & 1.28 & 0.821 & 2.33 & 0.768 & 6.10 & 0.802 \\
\hspace{3mm} \textbf{MM-Tokenizer} & Cascade & 1.16 & 0.832 & 1.74 & 0.782 & 6.14 & \textbf{0.817} \\
\hspace{3mm} \textbf{MM-Tokenizer} & E2E & \textbf{0.97} & \textbf{0.836} & 1.70 & \textbf{0.794} & 6.06 & 0.799 \\

\midrule 

\multicolumn{8}{l}{\textit{Frame Rate: 12.5Hz}} \\
\hspace{3mm} \textbf{MM-Tokenizer} & Cascade & 1.57 & 0.794 & 2.19 & 0.718 & 6.98 & 0.778 \\
\hspace{3mm} \textbf{MM-Tokenizer} & E2E & 1.01 & 0.828 & \textbf{1.63} & 0.785 & \textbf{6.07} & 0.798 \\

\bottomrule
\end{tabular}
\vspace{2mm}
\caption{\textbf{Zero-shot TTS performance on the SEED-TTS-Eval benchmark}. We compare the proposed MM-Tokenizer against the $S^3$-Tokenizer baseline~\cite{du2024cosyvoice} at different frame rates. \textbf{Bold} indicates the best result. ``Arch.'' denotes the generation architecture (Cascade vs. End-to-End) with the JoyVoice backbone. LLM decodes with \textbf{speaker embedding}.}
\label{tab:MM_tokenizer_downstream}
\end{table}

As outlined in Section~\ref{sec:mm_tokenizer}, the MM-Tokenizer is constructed by integrating a Finite Scalar Quantization (FSQ) layer and a CNN-based downsampling layer into the Whisper-large-v3 backbone. The model is optimized using a hybrid objective that combines supervised discriminative task losses with unsupervised reconstruction losses. Table~\ref{tab:MM_tokenizer_upstream} presents a comprehensive evaluation of the MM-Tokenizer across various audio understanding benchmarks, including Automatic Speech Recognition (ASR), Speech Emotion Recognition (SER), and Audio Event Detection (AED). The results demonstrate that the MM-Tokenizer (at both 25Hz and 12.5Hz frame rates) maintains competitive semantic retention compared to the Whisper-large-v3 baseline.

Table~\ref{tab:MM_tokenizer_downstream} presents the zero-shot TTS performance results on the SEED-TTS-Eval benchmark using the JoyVoice backbone. At a standard frame rate of 25Hz, the proposed \emph{MM-Tokenizer} consistently outperforms the baseline $S^3$-Tokenizer across the majority of metrics. Specifically, the End-to-End (E2E) variant of the MM-Tokenizer achieves superior intelligibility and similarity on the Chinese dataset (\textit{test-zh}), recording the lowest Character Error Rate (CER) of 0.97 and the highest Speaker Similarity (SS) of 0.836. While the Cascade architecture at 25Hz performs competitively achieving the highest similarity score on the \textit{test-hard} subset (0.817). The E2E model demonstrates remarkable robustness when the frame rate is reduced to 12.5Hz. Despite the lower temporal resolution, the 12.5Hz E2E model attains the state-of-the-art Word Error Rate (WER) of 1.63 on the English test set (\textit{test-en}) and the lowest CER of 6.07 on the challenging \textit{test-hard} set, validating the efficiency of the proposed tokenizer design.

\subsection{Single-Speaker Zero-shot Voice Clone}

\begin{table*}[htb]
	\centering
	\setlength\tabcolsep{4.5pt}
	\scalebox{0.80}{
		\begin{tabular}{lclclcl}
			\toprule
			\multirow{2}{*}{\textbf{Model}} & \multicolumn{2}{c}{\textbf{\emph{test-zh}}} & \multicolumn{2}{c}{\textbf{\emph{test-en}}} & \multicolumn{2}{c}{\textbf{\emph{test-hard}}} \\
			\cmidrule(r){2-3} \cmidrule(r){4-5} \cmidrule(r){6-7}
			& \textbf{CER (\%)~$\downarrow$} & \multicolumn{1}{c}{\textbf{SS~$\uparrow$}} & \textbf{WER (\%)~$\downarrow$} & \multicolumn{1}{c}{\textbf{SS~$\uparrow$}} & \textbf{CER (\%)~$\downarrow$} & \multicolumn{1}{c}{\textbf{SS~$\uparrow$}}  \\
			\midrule
			\textbf{Human} & 1.26 & 0.755~(0.775) & 2.14 & 0.734~(0.742)  & - & \multicolumn{1}{c}{-} \\
			\textbf{Vocoder Resyn.} & 1.27 & 0.720 & 2.17 & 0.700 & - & \multicolumn{1}{c}{-} \\
			\midrule
                \multicolumn{7}{c}{\textbf{Single-speaker Models}} \\
                \midrule
                \textbf{Seed-TTS}~\cite{anastassiou2024seed} & {1.12} & 0.796 & 2.25 & \textbf{0.762}  & 7.59 & 0.776 \\
                \textbf{MiniMax-Speech}~\cite{zhang2025minimax} & 0.99 & \textbf{0.799} & 1.90 & 0.738 &  - & - \\
			\textbf{MaskGCT}~\cite{wang2024maskgct} & 2.27 & 0.774~(0.752) & 2.62 & 0.714~(0.730)  & 10.27 & 0.748~(0.720) \\
			\textbf{F5-TTS (32 NFE)}~\cite{chen2025f5tts} & 1.56 & 0.741~(0.794) & 1.83 & 0.647~(0.742)  & 8.67 & 0.713~(0.762) \\
                \textbf{Spark TTS}~\cite{wang2025spark} & 1.20 & 0.672 & {1.98} & 0.584  & - & \multicolumn{1}{c}{-} \\
                \textbf{Qwen2.5-Omni-7B$_{RL}$}~\cite{xu2025qwen2} & 1.42 & 0.754 & 2.33 & 0.641  & 6.54 & 0.752 \\
                \textbf{CosyVoice 2}~\cite{du2024cosyvoice2} & 1.45 & 0.748~(0.806) & 2.57 & 0.652~(0.736)  &  6.83 & 0.724~(0.776) \\
                \textbf{CosyVoice 3-0.5B}~\cite{du2025cosyvoice3} & 1.16 & 0.780~(0.825) & 2.02 & 0.718~(0.789) & 6.08 & {0.758}~(\underline{0.815}) \\
                \textbf{CosyVoice 3-0.5B$_{{RL}}$}~\cite{du2025cosyvoice3} & \underline{0.75} & 0.774~(\underline{0.836}) & 1.76 & 0.695~(0.783) & \textbf{5.09} & 0.750~(0.809) \\
                \midrule
                \multicolumn{7}{c}{\textbf{Multi-speaker Models}} \\
                \midrule
                \textbf{Mooncast$^*$}~\cite{ju2025mooncast} & 6.72 & 0.550~(0.559) & 9.77 & 0.437~(0.463)  & \multicolumn{1}{c}{-} & \multicolumn{1}{c}{-} \\
                \textbf{MOSS-TTSD$^*$}~\cite{moss2025ttsd} & 3.05 & 0.678~(0.669) & 17.64 & 0.594~(0.628)  & 23.21 & 0.644~(0.633) \\
                \textbf{VibeVoice-1.5B$^*$}~\cite{peng2025vibevoice} & 2.86 & 0.689~(0.656) & 6.61 & 0.587~(0.595)  & 25.59 & 0.612~(0.587) \\
                \textbf{VibeVoice-7B$^*$}~\cite{peng2025vibevoice} & 1.60 & 0.707~(0.676) & 4.42 & 0.604~(0.594)  & 21.65 & 0.640~(0.618) \\
                
                \midrule
                \textbf{JoyVoice-Cascade} & 1.13 & 0.780~(0.825) & 1.75 & 0.710~(0.778) & 6.07 & 0.764~({0.805}) \\
                \textbf{JoyVoice-Cascade$_{{RL}}$} & \textbf{0.73} & 0.786~(0.831) & \textbf{1.45} & 0.718~(0.786) & 5.56 & 0.774~({0.813}) \\
                \textbf{JoyVoice} & 0.97 & 0.786~({0.827}) & {1.69} & 0.736~(0.790) & \underline{5.55} & 0.746~(0.773) \\
                \textbf{JoyVoice-Streaming} & 0.97 & 0.790~({0.831}) & {1.70} & \underline{0.738}~(\underline{0.793}) & 5.62 & \underline{0.777}~(0.809) \\
                \hspace{0.5em}+ llm spk embed. & 0.97 & \underline{0.797}~(\textbf{0.838}) & {1.69} & 0.737~(\textbf{0.795}) & 5.97 & \textbf{0.784}~(\textbf{0.820}) \\
                \textbf{JoyVoice-12.5Hz} & 0.95 & 0.778~({0.824}) & \underline{1.64} & 0.717~(0.783) & 5.90 & 0.745~(0.780) \\
			\bottomrule
	\end{tabular}}
    \caption{\textbf{Zero-shot TTS Performance Evaluation on SEED-TTS-Eval Benchmark.} 
    We compare JoyVoice against state-of-the-art baselines on content consistency (CER/WER $\downarrow$) and speaker similarity (SS $\uparrow$). 
    SS scores are reported in the format: \textit{WavLM} (\textit{ERes2Net}). 
    \textbf{Bold} and \underline{underline} denote the best and second-best results, respectively. 
    ``-'' indicates unavailable data. ``$^*$'' indicates the result decoding through the open-source model by ourself.}
	\label{tab:tts-seed-test}
\end{table*}

Table~\ref{tab:tts-seed-test} presents the TTS performance of JoyVoice and several recent models across the SEED test sets.
The evaluation focuses on content consistency (WER/CER) and speaker similarity (SS). The capability of Single-speaker Models is restricted to generating single-speaker audio. Multi-speaker Models, on the other hand, possess the functionality to produce long-form, interactive audio involving multiple speakers.

Although designed as a versatile model capable of multi-speaker, long-form audio synthesis, JoyVoice also achieves state-of-the-art performance on the single-speaker voice cloning benchmark.
JoyVoice achieves significant improvements over JoyVoice-Cascade, with relative gains of 14.2\% (1.13\% $\to$ 0.97\%) on \emph{test-zh} and 3.4\% (1.75\% $\to$ 1.69\%) on \emph{test-en}. In the \emph{test-hard} set, JoyVoice reduces the CER from 6.07\% to 5.55\% (8.6\% relative improvement). Compared to other baselines w/o RL, JoyVoice consistently excels across all metrics.
Notably, application of our \textbf{RL policy} to the JoyVoice-Cascade model deliveres relative CER reductions of 48.4\%, 17.2\%, and 8.4\% on the \emph{test-zh},  \emph{test-en}, and \emph{test-hard}, respectively.

Regarding \textbf{streaming}, the JoyVoice served as the initialization, wherein the LLM component was kept frozen, and training was conducted exclusively on the Dynamic Chunk Flow Matching module.
As evidenced by the results in the table~\ref{tab:tts-seed-test}, JoyVoice-Streaming not only maintains performance parity with JoyVoice in terms of content consistency but also achieves a substantial improvement in speaker similarity on the challenging \emph{test-hard} set.
By adopting a dynamic chunk-based training strategy, our JoyVoice-Streaming permits the use of variable chunk sizes during inference, thereby enabling precise control over computational latency to meet diverse application demands. The associated performance metrics are compiled in table~\ref{tab:llm_spk}.

Furthermore, an ablation study in table~\ref{tab:llm_spk} compares the performance of the LLM component with and without speaker embedding. The results consistently demonstrate that incorporating speaker embedding degrades the CER/WER performance, but nevertheless leads to a substantial improvement in speaker similarity. Moreover, our JoyVoice-Streaming with speaker embedding at chunk size 24/48 surpasses CosyVoice 3-0.5B, one of the current top-performing streaming models, in both content consistency and speaker similarity.


\begin{table}[htb]
\centering
\renewcommand{\arraystretch}{1.2} 
\setlength\tabcolsep{5pt} 

\begin{tabular}{lccccccc} 
\toprule
\multirow{2.5}{*}{\textbf{Model}} & \multirow{2.5}{*}{\textbf{Chunk}} & \multicolumn{2}{c}{\textbf{\textit{test-zh}}} & \multicolumn{2}{c}{\textbf{\textit{test-en}}} & \multicolumn{2}{c}{\textbf{\textit{test-hard}}} \\
\cmidrule(lr){3-4} \cmidrule(lr){5-6} \cmidrule(lr){7-8} 
 & & \textbf{CER}$\downarrow$ & \textbf{SS}$\uparrow$ & \textbf{WER}$\downarrow$ & \textbf{SS}$\uparrow$ & \textbf{CER}$\downarrow$ & \textbf{SS}$\uparrow$ \\
\midrule

\textbf{CosyVoice 3-0.5B} & full & 1.16 & 0.825 & 2.02 & 0.789 & 6.08 & 0.815 \\
\textbf{JoyVoice} & full & 0.97 & 0.827 & \textbf{1.69} & 0.790 & \textbf{5.55} & 0.773 \\
\hspace{0.5em}+ llm spk embed. & full & 0.97 & 0.836 & 1.70 & 0.794 & 6.06 & 0.799 \\
\textbf{JoyVoice-Streaming} & 48 & 0.97 & 0.831 & 1.70 & 0.793 & 5.62 & 0.809 \\
\hspace{0.5em}+ llm spk embed. & 48 & 0.97 & \textbf{0.838} & \textbf{1.69} & \textbf{0.795} & 5.97 & \textbf{0.820} \\
\textbf{JoyVoice-Streaming} & 24 & 0.98 & 0.831 & 1.70 & 0.793 & 5.69 & 0.809 \\
\hspace{0.5em}+ llm spk embed. & 24 & \textbf{0.96} & \textbf{0.838} & 1.73 & 0.794 & 5.97 & \textbf{0.820} \\

\bottomrule
\end{tabular}
\vspace{2mm}
\caption{Evaluation of zero-shot TTS performance on the SEED-TTS-Eval dataset, comparing the performance of the LLM component with different chunk size. Bold numbers indicate the best results in each column.}
\label{tab:llm_spk}
\end{table}

\subsection{Multi-Speaker Zero-shot Voice Clone}

Table~\ref{tab:msmt-res} presents the multi-speaker, long-form audio synthesis performance of JoyVoice and several recent models across the our JoyVoice-MSMT-eval test sets, which include the Chinese \emph{test-zh-msmt} and English \emph{test-en-msmt}. The evaluation not only focuses on CER/WER, but also  and focuses on cpCER/cpWER. 
By achieving the lowest WER and cpWER on the 2-4 speaker MSMT test sets, these results underscore that JoyVoice possesses an exceptional capability to preserve both content and speaker timbre faithfully.
It is observed that the cpWER for 4-speaker results exhibits a noticeable degradation compared to the 2- and 3-speaker results. This can be primarily attributed to the under-representation of conversations with more than four speakers in our training data distribution, as detailed in Figure~\ref{fig:spk_dist}. To address this limitation, augmenting our dataset with a higher proportion of multi-speaker samples will be prioritized in future work to enhance performance in these settings.

For speaker identification, we rely on the open-source model pyannote~\cite{bredin23_interspeech}, and the corresponding metrics should be interpreted with this context in mind, as they reflect the capabilities and limitations of this external system. Our subjective listening test website offers a complementary perspective for a comprehensive assessment.

\begin{table}[htb] 
\centering
\renewcommand{\arraystretch}{1.25} 
\setlength\tabcolsep{6pt} 

\resizebox{\textwidth}{!}{
\begin{tabular}{lcccccc} 
\toprule
\multirow{3}{*}{\textbf{Model}} & \multicolumn{3}{c}{\textbf{\textit{test-zh-msmt}}} & \multicolumn{3}{c}{\textbf{\textit{test-en-msmt}}} \\
& \multicolumn{3}{c}{\small (CER / cpCER) $\downarrow$} & \multicolumn{3}{c}{\small (WER / cpWER) $\downarrow$} \\
\cmidrule(lr){2-4} \cmidrule(lr){5-7} 
 & \textbf{2spk} & \textbf{3spk} & \textbf{4spk} & \textbf{2spk} & \textbf{3spk} & \textbf{4spk} \\
\midrule
\textbf{Mooncast$^*$}      & 7.57 / 12.62 & - & - & 5.32 / 17.38 & - & - \\
\textbf{MOSS-TTSD$^*$}     & 3.00 / 6.02  & - & - & 5.56 / 10.04 & - & - \\
\textbf{VibeVoice-1.5B$^*$} & 5.54 / 14.88 & 6.30 / 16.77 & 5.29 / 17.20 & 4.26 / 7.73  & 8.56 / 12.22 & 8.89 / 14.11 \\
\textbf{VibeVoice-7B$^*$}   & 1.80 / 7.57  & 3.15 / 9.37  & 3.66 / 16.60 & 4.19 / 6.58  & 6.66 / 10.60 & 6.42 / 13.81 \\
\textbf{JoyVoice}          & \textbf{1.44} / \textbf{1.88} & \textbf{1.83} / \textbf{4.94} & \textbf{2.08} / \textbf{13.34} & \textbf{3.36} / \textbf{3.61} & \textbf{6.01} / \textbf{9.57} & \textbf{3.93} / \textbf{12.18} \\
\bottomrule
\end{tabular}
}
\vspace{2mm}
\caption{\textbf{Multi-speaker TTS model performance Evaluation on JoyVoice-MSMT-eval Benchmarks.} We compare JoyVoice against state-of-the-art baselines. The results are reported in the format of \textbf{CER/cpCER} for Chinese and \textbf{WER/cpWER} for English datasets (lower is better). \textbf{Boldface} denotes the best result. ``$^*$'' indicates results decoded using the open-source model by ourselves.}
\label{tab:msmt-res}
\end{table}

\subsection{Speaker Fine-tuned Models}

To investigate the adaptation of a pre-trained foundation model to multiple target speakers, we employed the multi-speaker fine-tuning (mSFT) paradigm as introduced in CosyVoice2. The objective was to derive distinct, high-quality voice profiles from a single model. Specifically, we selected five speakers (designated as Speaker~1 through Speaker~5) for this study. Each speaker's identity was explicitly conditioned during training by incorporating a speaker prompt tag (e.g., ``You are Speaker~1'') into the system prompt. The total duration of the speech data used for fine-tuning was approximately 12 hours, with individual speakers contributing between 2 to 3 hours of audio.

\begin{figure*}[thb]
	\centering
	\includegraphics[width=\linewidth]{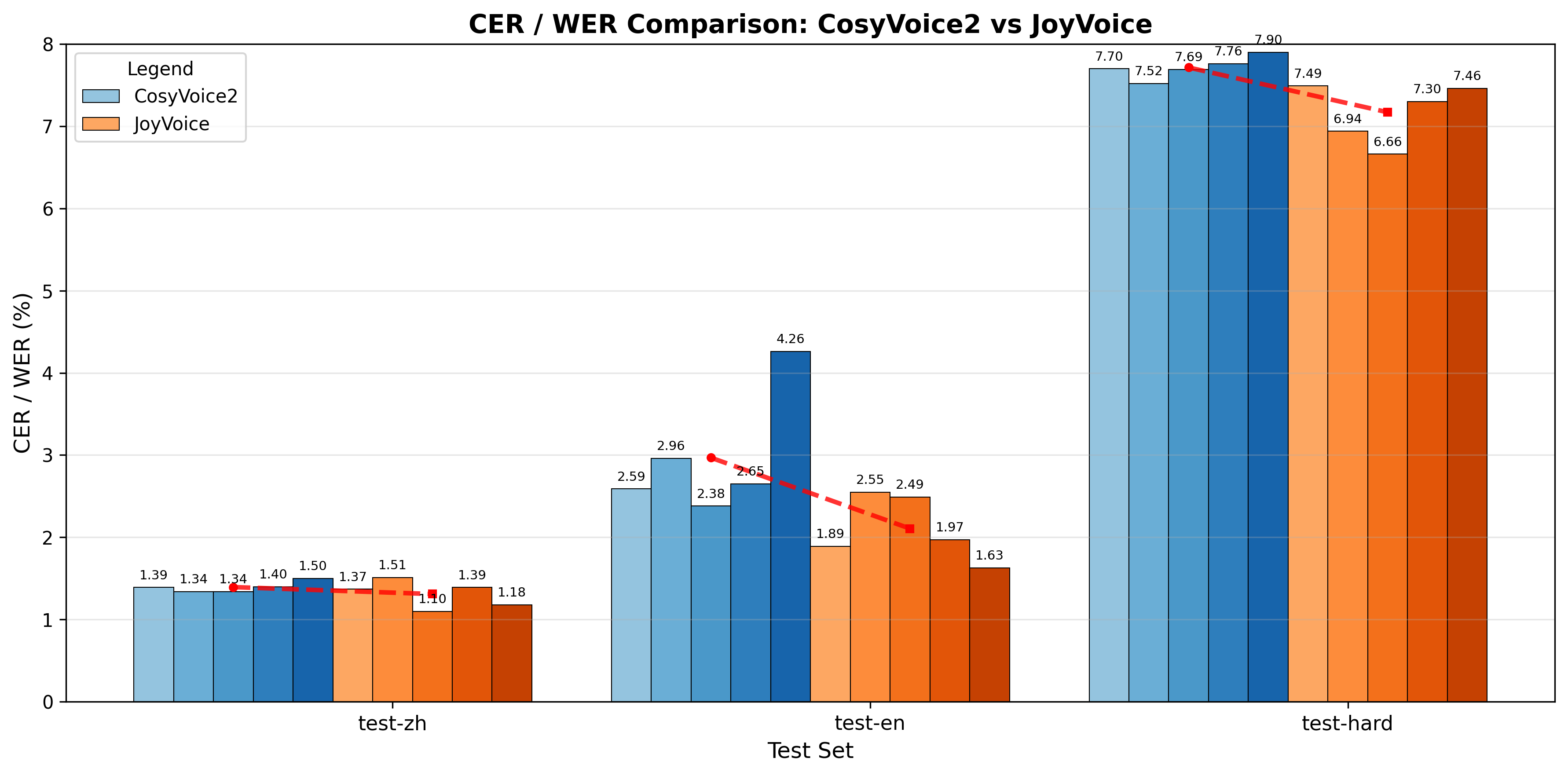}
	\caption{Comparison of full-parameter SFT results on five speakers.}
	\label{fig:sft}
\end{figure*}

For the mSFT process, we maintained a consistent learning rate of $1\times10^{-5}$. A key aspect of our investigation was to compare the efficacy and efficiency of different fine-tuning strategies. To this end, we experimented with both full-parameter SFT and the parameter-efficient Low-Rank Adaptation (LoRA-SFT) method. Our findings indicate that the choice of strategy should be data-dependent. Based on our empirical observations, we recommend utilizing LoRA-SFT when the target speaker's data is limited (e.g., less than 1 hour) to effectively mitigate the risk of overfitting. A detailed comparison of the results from these approaches is presented in Figure~\ref{fig:sft}.

Traditional Supervised Fine-Tuning (SFT) for TTS involves training individual models using single-speaker data. Since the JoyVoice foundation model supports multi-speaker generation, we innovated by fine-tuning it with approximately one hour of two-speaker conversational podcast data. This yielded a sophisticated dual-speaker TTS model that accurately captures the timbre, speaking style, and prosodic naturalness of the source voices. For subjective listening test results, please refer to the demo page.

%% file: chapters/4_conclusion.tex
\section{Conclusion}

In this work, we introduced JoyVoice, a novel and powerful foundation model for conversational speech synthesis. Designed to overcome the limitations of prior models, JoyVoice is a highly anthropomorphic, multi-speaker, and long-context system that enables realistic dialogue generation. Demonstrating its advanced capability, the model can synthesize natural conversations featuring up to eight distinct speakers and extending up to five minutes in duration within a single generation shot. The superior performance of JoyVoice is rooted in several key technical advancements. We proposed an innovative End-to-End Transformer-DiT Architecture specifically optimized for managing complex conversational dynamics and long-term context. Complementing this architecture, we introduced the novel MM-Tokenizer audio discretization scheme. Furthermore, through strategic application of large-scale data and extensive data augmentation, we successfully developed a robust audio generation model that significantly reduces or eliminates the dependency on traditional, often fragile, text pre-processing pipelines.  Comprehensive evaluations confirm that JoyVoice achieves substantial improvements over similar speech foundation models.  Specifically, JoyVoice achieves top-tier results on both the Seed-TTS-Eval Benchmark and multi-speaker long-form conversational voice cloning tasks, demonstrating superior audio quality and generalization.

%% file: chapters/5_limitation.tex
\section{Limitations}
While JoyVoice achieves state-of-the-art performance in highly anthropomorphic, multi-speaker conversational speech synthesis, several inherent limitations remain and motivate our ongoing research.

First, the stability and quality of the synthesized audio currently show degradation when generating conversations involving more than four speakers. We attribute this primary issue to the insufficient coverage of highly dynamic, natural multi-speaker interaction data within our current training corpus. Future work will prioritize the acquisition of more extensive and complex conversational datasets to ensure robustness and high fidelity across increasingly complex dialogue scenarios.

Second, the large-scale Reinforcement Learning (RL) training phase for JoyVoice is still in progress. Beyond conventional objectives such as optimizing text consistency and speaker timbre similarity in the synthesized output, we plan to further explore the utility of RL. Specifically, we aim to investigate how RL can be effectively utilized to boost the long-term stability of multi-speaker synthesis and to significantly improve the model's ability to express anthropomorphic emotions and nuanced sentiment over extended conversational contexts.

Finally, JoyVoice is currently restricted to pure speech generation and does not yet support the synthesis of general audio events or music. Achieving a unified, large-scale audio generation model that encompasses speech, audio, and music demands substantial, in-depth exploration into two main technical areas: the design of a universal audio tokenizer and the development of a highly generalized model architecture capable of modeling diverse audio modalities.

\section*{Authors in Alphabetical Order}

\begin{description}
    \item[Core Contributors:] 
    Fan Yu, Tao Wang, You Wu, Lin Zhu, Wei Deng, Weisheng Han, Wenchao Wang, Lin Hu, Xiangyu Liang, Xiaodong He, Yankun Huang, Yu Gu, Yuan Liu, Yuxuan Wang, Zhangyu Xiao, Ziteng Wang
    
    \item[Contributors:] 
   Boya Dong, Feng Dang, Jinming Chen, Jingdong Li, Jun Wang, Yechen Jin, Yuan Zhang, Zhengyan Sheng, Xin Wang
   
\end{description}

%% file: ref.bib
@inproceedings{radford2023robust,
  title={Robust speech recognition via large-scale weak supervision},
  author={Radford, Alec and Kim, Jong Wook and Xu, Tao and Brockman, Greg and McLeavey, Christine and Sutskever, Ilya},
  booktitle={International conference on machine learning},
  pages={28492--28518},
  year={2023},
  organization={PMLR}
}

@article{mentzer2023finite,
  title={Finite scalar quantization: Vq-vae made simple},
  author={Mentzer, Fabian and Minnen, David and Agustsson, Eirikur and Tschannen, Michael},
  journal={arXiv preprint arXiv:2309.15505},
  year={2023}
}

@article{du2024cosyvoice,
  title={Cosyvoice: A scalable multilingual zero-shot text-to-speech synthesizer based on supervised semantic tokens},
  author={Du, Zhihao and Chen, Qian and Zhang, Shiliang and Hu, Kai and Lu, Heng and Yang, Yexin and Hu, Hangrui and Zheng, Siqi and Gu, Yue and Ma, Ziyang and others},
  journal={arXiv preprint arXiv:2407.05407},
  year={2024}
}

@article{du2024cosyvoice2,
  title={Cosyvoice 2: Scalable streaming speech synthesis with large language models},
  author={Du, Zhihao and Wang, Yuxuan and Chen, Qian and Shi, Xian and Lv, Xiang and Zhao, Tianyu and Gao, Zhifu and Yang, Yexin and Gao, Changfeng and Wang, Hui and others},
  journal={arXiv preprint arXiv:2412.10117},
  year={2024}
}

@article{du2025cosyvoice3,
  title={Cosyvoice 3: Towards in-the-wild speech generation via scaling-up and post-training},
  author={Du, Zhihao and Gao, Changfeng and Wang, Yuxuan and Yu, Fan and Zhao, Tianyu and Wang, Hao and Lv, Xiang and Wang, Hui and Ni, Chongjia and Shi, Xian and others},
  journal={arXiv preprint arXiv:2505.17589},
  year={2025}
}

@inproceedings{chen2025f5tts,
  title={{F5-TTS}: A fairytaler that fakes fluent and faithful speech with flow matching},
  author={Chen, Yushen and Niu, Zhikang and Ma, Ziyang and Deng, Keqi and Wang, Chunhui and JianZhao, JianZhao and Yu, Kai and Chen, Xie},
  booktitle={Proceedings of the 63rd Annual Meeting of the Association for Computational Linguistics (Volume 1: Long Papers)},
  pages={6255--6271},
  year={2025}
}

@article{vall-e,
  title={Neural codec language models are zero-shot text to speech synthesizers},
  author={Chen, Sanyuan and Wang, Chengyi and Wu, Yu and Zhang, Ziqiang and Zhou, Long and Liu, Shujie and Chen, Zhuo and Liu, Yanqing and Wang, Huaming and Li, Jinyu and others},
  journal={IEEE Transactions on Audio, Speech and Language Processing},
  year={2025},
  publisher={IEEE}
}

@article{ye2025llasa,
  title={Llasa: Scaling Train-Time and Inference-Time Compute for {Llama}-based Speech Synthesis},
  author={Ye, Zhen and Zhu, Xinfa and Chan, Chi-Min and Wang, Xinsheng and Tan, Xu and Lei, Jiahe and Peng, Yi and Liu, Haohe and Jin, Yizhu and DAI, Zheqi and others},
  journal={arXiv e-prints},
  pages={arXiv--2502},
  year={2025}
}

@article{zhu2025zipvoice,
  title={ZipVoice: Fast and High-Quality Zero-Shot Text-to-Speech with Flow Matching},
  author={Zhu, Han and Kang, Wei and Yao, Zengwei and Guo, Liyong and Kuang, Fangjun and Li, Zhaoqing and Zhuang, Weiji and Lin, Long and Povey, Daniel},
  journal={arXiv preprint arXiv:2506.13053},
  year={2025}
}

@article{lipman2022flowmatching,
  title={Flow matching for generative modeling},
  author={Lipman, Yaron and Chen, Ricky TQ and Ben-Hamu, Heli and Nickel, Maximilian and Le, Matt},
  journal={arXiv preprint arXiv:2210.02747},
  year={2022}
}

@article{guo2024fireredtts,
  title={FireRedTTS: A Foundation Text-To-Speech Framework for Industry-Level Generative Speech Applications},
  author={Guo, Hao-Han and Liu, Kun and Shen, Fei-Yu and Wu, Yi-Chen and Xie, Feng-Long and Xie, Kun and Xu, Kai-Tuo},
  journal={arXiv e-prints},
  pages={arXiv--2409},
  year={2024}
}

@article{anastassiou2024seed,
  title={Seed-tts: A family of high-quality versatile speech generation models},
  author={Anastassiou, Philip and Chen, Jiawei and Chen, Jitong and Chen, Yuanzhe and Chen, Zhuo and Chen, Ziyi and Cong, Jian and Deng, Lelai and Ding, Chuang and Gao, Lu and others},
  journal={arXiv preprint arXiv:2406.02430},
  year={2024}
}

@article{peng2025vibevoice,
  title={Vibevoice technical report},
  author={Peng, Zhiliang and Yu, Jianwei and Wang, Wenhui and Chang, Yaoyao and Sun, Yutao and Dong, Li and Zhu, Yi and Xu, Weijiang and Bao, Hangbo and Wang, Zehua and others},
  journal={arXiv preprint arXiv:2508.19205},
  year={2025}
}

@article{jia2025ditar,
  title={Ditar: Diffusion transformer autoregressive modeling for speech generation},
  author={Jia, Dongya and Chen, Zhuo and Chen, Jiawei and Du, Chenpeng and Wu, Jian and Cong, Jian and Zhuang, Xiaobin and Li, Chumin and Wei, Zhen and Wang, Yuping and others},
  journal={arXiv preprint arXiv:2502.03930},
  year={2025}
}

@inproceedings{peebles2023dit,
  title={Scalable diffusion models with transformers},
  author={Peebles, William and Xie, Saining},
  booktitle={Proceedings of the IEEE/CVF international conference on computer vision},
  pages={4195--4205},
  year={2023}
}

@article{rafailov2023direct,
  title={Direct preference optimization: Your language model is secretly a reward model},
  author={Rafailov, Rafael and Sharma, Archit and Mitchell, Eric and Manning, Christopher D and Ermon, Stefano and Finn, Chelsea},
  journal={Advances in neural information processing systems},
  volume={36},
  pages={53728--53741},
  year={2023}
}

@article{liu2024ardit,
  title={Autoregressive diffusion transformer for text-to-speech synthesis},
  author={Liu, Zhijun and Wang, Shuai and Inoue, Sho and Bai, Qibing and Li, Haizhou},
  journal={arXiv preprint arXiv:2406.05551},
  year={2024}
}

@article{wang2025spark,
  title={Spark-tts: An efficient llm-based text-to-speech model with single-stream decoupled speech tokens},
  author={Wang, Xinsheng and Jiang, Mingqi and Ma, Ziyang and Zhang, Ziyu and Liu, Songxiang and Li, Linqin and Liang, Zheng and Zheng, Qixi and Wang, Rui and Feng, Xiaoqin and others},
  journal={arXiv preprint arXiv:2503.01710},
  year={2025}
}

@article{zhou2025indextts2,
  title={IndexTTS2: A Breakthrough in Emotionally Expressive and Duration-Controlled Auto-Regressive Zero-Shot Text-to-Speech},
  author={Zhou, Siyi and Zhou, Yiquan and He, Yi and Zhou, Xun and Wang, Jinchao and Deng, Wei and Shu, Jingchen},
  journal={arXiv preprint arXiv:2506.21619},
  year={2025}
}

@article{zhang2025minimax,
  title={Minimax-speech: Intrinsic zero-shot text-to-speech with a learnable speaker encoder},
  author={Zhang, Bowen and Guo, Congchao and Yang, Geng and Yu, Hang and Zhang, Haozhe and Lei, Heidi and Mai, Jialong and Yan, Junjie and Yang, Kaiyue and Yang, Mingqi and others},
  journal={arXiv preprint arXiv:2505.07916},
  year={2025}
}

@article{bai2023qwen,
  title={Qwen technical report},
  author={Bai, Jinze and Bai, Shuai and Chu, Yunfei and Cui, Zeyu and Dang, Kai and Deng, Xiaodong and Fan, Yang and Ge, Wenbin and Han, Yu and Huang, Fei and others},
  journal={arXiv preprint arXiv:2309.16609},
  year={2023}
}

@article{qwen2.5,
    title   = {Qwen2.5 Technical Report}, 
    author  = {An Yang and Baosong Yang and Beichen Zhang and Binyuan Hui and Bo Zheng and Bowen Yu and Chengyuan Li and Dayiheng Liu and Fei Huang and Haoran Wei and Huan Lin and Jian Yang and Jianhong Tu and Jianwei Zhang and Jianxin Yang and Jiaxi Yang and Jingren Zhou and Junyang Lin and Kai Dang and Keming Lu and Keqin Bao and Kexin Yang and Le Yu and Mei Li and Mingfeng Xue and Pei Zhang and Qin Zhu and Rui Men and Runji Lin and Tianhao Li and Tingyu Xia and Xingzhang Ren and Xuancheng Ren and Yang Fan and Yang Su and Yichang Zhang and Yu Wan and Yuqiong Liu and Zeyu Cui and Zhenru Zhang and Zihan Qiu},
    journal = {arXiv preprint arXiv:2412.15115},
    year    = {2024}
}

@article{ju2025mooncast,
  title={{MoonCast}: High-quality zero-shot podcast generation},
  author={Ju, Zeqian and Yang, Dongchao and Yu, Jianwei and Shen, Kai and Leng, Yichong and Wang, Zhengtao and Tan, Xu and Zhou, Xinyu and Qin, Tao and Li, Xiangyang},
  journal={arXiv preprint arXiv:2503.14345},
  year={2025}
}

@article{wang2024maskgct,
  title={Maskgct: Zero-shot text-to-speech with masked generative codec transformer},
  author={Wang, Yuancheng and Zhan, Haoyue and Liu, Liwei and Zeng, Ruihong and Guo, Haotian and Zheng, Jiachen and Zhang, Qiang and Zhang, Xueyao and Zhang, Shunsi and Wu, Zhizheng},
  journal={arXiv preprint arXiv:2409.00750},
  year={2024}
}

@article{xu2025qwen2,
  title={Qwen2. 5-omni technical report},
  author={Xu, Jin and Guo, Zhifang and He, Jinzheng and Hu, Hangrui and He, Ting and Bai, Shuai and Chen, Keqin and Wang, Jialin and Fan, Yang and Dang, Kai and others},
  journal={arXiv preprint arXiv:2503.20215},
  year={2025}
}

@article{moss2025ttsd,
  title={Text to Spoken Dialogue Generation},
  author={OpenMOSS},
  year={2025}
}

@inproceedings{Yu2022ACS,
  title={A Comparative study on speaker-attributed automatic speech recognition in multi-party meetings},
  author={Fan Yu and Zhihao Du and ShiLiang Zhang and Yuxiao Lin and Lei Xie},
  year={2022},
  booktitle={Proc. INTERSPEECH},
  pages={560--564},
  organization={ISCA}
}

@inproceedings{kanda2021comparative,
  title={A comparative study of modular and joint approaches for speaker-attributed {ASR} on monaural long-form audio},
  author={Kanda, Naoyuki and Xiao, Xiong and Wu, Jian and Zhou, Tianyan and Gaur, Yashesh and Wang, Xiaofei and Meng, Zhong and Chen, Zhuo and Yoshioka, Takuya},
  booktitle={Proc. ASRU},
  pages={296--303},
  year={2021},
  organization={IEEE}
}

@article{zhu2025zipvoicedialognonautoregressivespokendialogue,
  title={ZipVoice-Dialog: Non-Autoregressive Spoken Dialogue Generation with Flow Matching}, 
  author={Han Zhu and Wei Kang and Liyong Guo and Zengwei Yao and Fangjun Kuang and Weiji Zhuang and Zhaoqing Li and Zhifeng Han and Dong Zhang and Xin Zhang and Xingchen Song and Long Lin and Daniel Povey},
  journal={arXiv preprint arXiv:2507.09318},
  year={2025}
}

@inproceedings{bredin23_interspeech,
  title     = {pyannote.audio 2.1 speaker diarization pipeline: principle, benchmark, and recipe},
  author    = {Hervé Bredin},
  year      = {2023},
  booktitle = {Interspeech 2023},
  pages     = {1983--1987},
  doi       = {10.21437/Interspeech.2023-105},
  issn      = {2958-1796},
}
